# Scalable readout interface for superconducting nanowire single-photon detectors using AQFP and RSFQ logic families

NAOKI TAKEUCHI,[1,*] FUMIHIRO CHINA,[2] SHIGEHITO MIKI,[2,3] SHIGEYUKI MIYAJIMA,[2] MASAHIRO YABUNO,[2] NOBUYUKI YOSHIKAWA,[1,4] AND HIROTAKA TERAI[2]

[1] *Institute of Advanced Sciences, Yokohama National University, 79-5 Tokiwadai, Hodogaya, Yokohama 240-8501, Japan*
[2] *Advanced ICT Research Institute, National Institute of Information and Communications Technology, 588-2 Iwaoka, Nishi, Kobe 651-2492, Japan*
[3] *Graduate School of Engineering Faculty of Engineering, Kobe University, 1-1 Rokkodai-cho, Nada, Kobe 657-0013, Japan*
[4] *Department of Electrical and Computer Engineering, Yokohama National University, 79-5 Tokiwadai, Hodogaya, Yokohama 240-8501, Japan*
*\* takeuchi-naoki-kx@ynu.ac.jp*

**Abstract:** We propose a scalable readout interface for superconducting nanowire single-photon detector (SSPD) arrays, which we call the AQFP/RSFQ interface. This interface is composed of adiabatic quantum-flux-parametron (AQFP) and rapid single-flux-quantum (RSFQ) logic families. The AQFP part reads out the spatial information of an SSPD array via a single cable, and the RSFQ part reads out the temporal information via a single cable. The hybrid interface has high temporal resolution owing to low timing jitter in the operation of the RSFQ part. In addition, the hybrid interface achieves high circuit scalability because of low supply current in the operation of the AQFP part. Therefore, the hybrid interface is suitable for handling many-pixel SSPD arrays. We demonstrate a four-pixel SSPD array using the hybrid interface as proof of concept. The measurement results show that the hybrid interface can read out all of the pixels with a low error rate and low timing jitter.

## 1. Introduction

Superconducting nanowire single-photon detectors (SSPDs or SNSPDs) [1] have superior performance in terms of detection efficiency, count rate, and timing jitter [2–4], and thus have been used in many research fields, such as quantum optics [5], quantum information [6], optical communications [7], and fluorescent correlation spectroscopy [8]. Lately, a great deal of effort has been spent on the development of multi-pixel SSPD arrays [9–12] in order to achieve single-photon imagers, which pave the way for new applications such as time-of-flight imaging [13], dark matter detection [14], quantum information processing with single photons [15], and integrated spectrometry at the single photon level [16]. In order to develop large single-photon imagers for the above applications, it is crucial to establish efficient readout schemes by which the SSPD pixels are read out via a few coaxial cables. This is because the number of available cables for demonstrating SSPDs is limited by the cooling power of the cryocooler. For instance, in our experimental setup using a 0.1-W Gifford–McMahon (GM) cryocooler [17], the number of available cables is less than approximately twenty, beyond which the large heat load increases the temperature of the sample stage significantly. To date, several efficient readout schemes, such as row-column multiplexing [18], frequency multiplexing [19], amplitude multiplexing [20], and the delay-line approach [21], have been reported.

We have been developing readout interfaces [11,22] for SSPD arrays using rapid single-flux-quantum (RSFQ) logic [23], which read out the SSPD pixels via a single cable by digitizing and encoding the spatiotemporal information of an SSPD array at cryogenic temperature. RSFQ logic is of sufficiently low power (~400 nW per Josephson junction) so as to be suitable as readout interfaces for SSPDs. However, the amount of bias current (~150 µA per Josephson junction) for an RSFQ interface increases with the circuit complexity. Therefore, at some point, large bias current and the parasitic resistance in the equipment (e.g., cables, connectors, and bonding wires) inside the cryocooler generate significant Joule heating and increase the temperature of the sample stage. We found that the temperature of the RSFQ interface chip placed on the sample stage in a 0.1-W GM cryocooler increased from 2.7 K to approximately 6 K by supplying a bias current of 370 mA to the RSFQ chip [24], which indicates that the scalability of RSFQ interfaces (i.e., the number of pixels that an RSFQ interface can handle) is limited by the amount of bias current. Therefore, readout interfaces that operate with both low power and low supply current are required in order to demonstrate many-pixel SSPD arrays.

In a previous study [25], we proposed using adiabatic quantum-flux-parametron (AQFP) logic [26] as readout interfaces for SSPDs because AQFP circuits can operate with both low power and low supply current. AQFP logic is a low-power superconductor logic family based on the quantum flux parametron (QFP) [27,28]. AQFP gates maximize the benefit of adiabatic switching [29,30] by reducing the characteristic times of Josephson junctions, thereby achieving a power dissipation of 7 pW per Josephson junction at 5-GHz operation [31]. Moreover, since AQFP gates are serially biased by AC excitation currents [32], the amount of the supply currents (a few milliamps) does not increase with the circuit complexity [33]. We developed an AQFP interface for a single-pixel SSPD and demonstrated photon detection with a low error rate in a 0.1-W GM cryocooler [34].

In the present study, we propose a scalable readout interface for multi-pixel SSPD arrays using AQFP logic, which we call the AQFP/RSFQ hybrid interface. This interface is composed of AQFP and RSFQ circuits. The former outputs the spatial information (which pixel absorbs a photon) via a single cable, and the latter outputs the temporal information (when a photon is absorbed) via a single cable. A drawback of AQFP logic is that its temporal resolution is low due to synchronous operation (i.e., signals are sampled by AC excitation currents). Thus, in the hybrid interface, temporal information is obtained by the RSFQ circuits, which have high temporal resolution owing to asynchronous operation with low timing jitter [11]. Furthermore, since most of the parts of the hybrid interface are designed using AQFP logic, the supply current for the hybrid interface does not increase with the circuit complexity. Therefore, the hybrid interface achieves both high temporal resolution and high circuit scalability and thus is suitable for handling many-pixel SSPD arrays. Here, we demonstrate a four-pixel SSPD array using a hybrid interface as proof of concept. The measurement results show that the hybrid interface can read out the SSPD pixels with a low error rate and low timing jitter.

## 2. AQFP/RSFQ hybrid interface

Figure 1(a) illustrates a block diagram of an AQFP/RSFQ hybrid interface for $N = 4$, where $N$ is the number of the pixels in the SSPD array. For simplicity, it is assumed that each pixel in the array is read out via an individual wire (i.e., the SSPD array is connected to the hybrid interface via $N$ wires) and that pulsed current $I_{in}$ appears on one of the $N$ wires when the SSPD array detects a photon. The blocks with blue frames represent AQFP circuits, and those with red frames represent RSFQ circuits. The AQFP part digitizes and encodes the signal currents from the SSPD array, generating voltage signals $V_{aqfp}$ for spatial information. The RSFQ part merges the signal currents using a stack of $N$ DC superconducting quantum interference devices (DC-SQUIDs) and generates voltage signals $V_{rsfq}$ for temporal information. Figure 1(b) shows typical waveforms for $N = 4$, where $I_{in1}$ through $I_{in4}$ are the signal currents from pixels 1 through 4 in the SSPD array, respectively. The rising edge of $V_{rsfq}$ shows the temporal information owing to event-driven asynchronous operation in the RSFQ part. The serial data on $V_{aqfp}$ show the spatial information (i.e., the pixel address of the

SSPD array), where $V_{aqfp}$ is in the form of unipolar return-to-zero encoding and the first bit represents a flag. The point is that the spatiotemporal information ($V_{rsfq}$ and $V_{aqfp}$) of the SSPD array can be read out using only two coaxial cables, which helps reduce the number of cables required for demonstrating an SSPD array. Furthermore, the amount of the supply current for the hybrid interface does not increase with $N$. While the complexity of the AQFP part increases with $N$, the amplitude of the excitation current for the AQFP part is independent of $N$ because of serial biasing. For the RSFQ part, only the number of DC-SQUIDs in the stack increases with $N$, which does not increase the amount of the bias current for the RSFQ part. Thus, the amount of the supply currents for both AQFP and RSFQ parts is independent of $N$, which indicates that the hybrid interface is suitable for handling many-pixel SSPD arrays with large $N$. The details of both AQFP and RSFQ parts are given below.

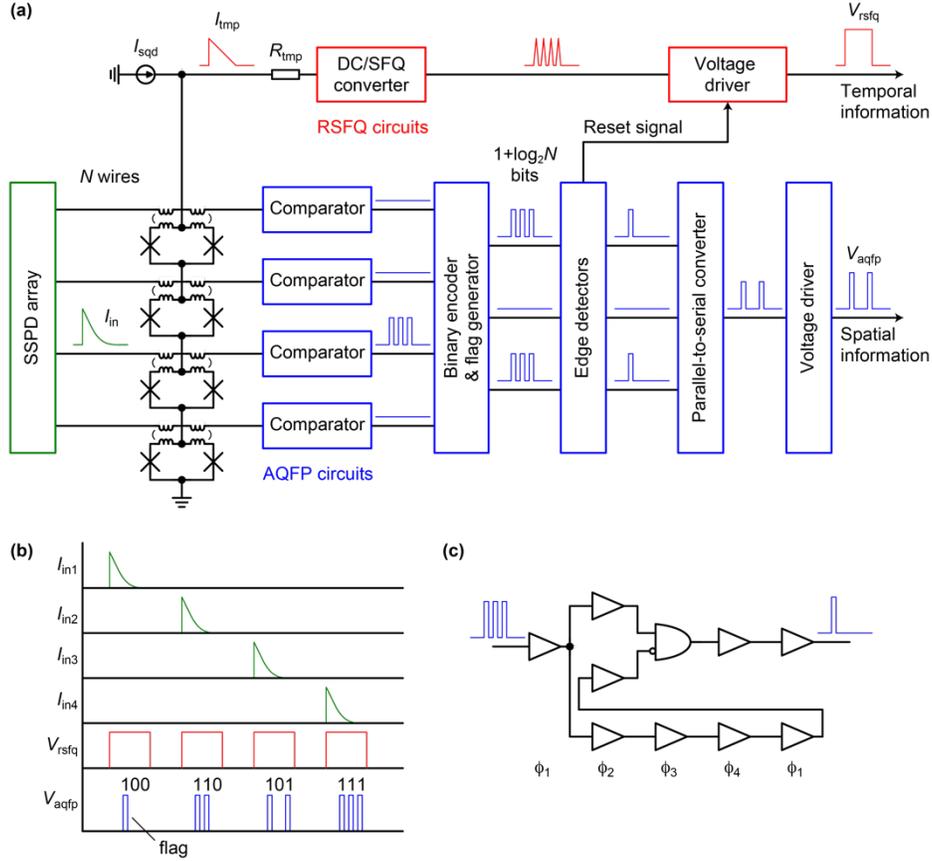

Fig. 1. AQFP/RSFQ hybrid interface. (a) Block diagram and (b) typical waveforms for $N = 4$. The rising edge of $V_{rsfq}$ represents the temporal information, and the serial data on $V_{aqfp}$ represent the spatial information. (c) Rising-edge detector, which converts a logic-1 train into a single logic 1.

The AQFP part operates as follows. Each wire from the SSPD array is terminated by a comparator [25]. The comparators sample $I_{in}$ in synchronization with AC excitation currents, thereby converting $I_{in}$ into a logic-1 train. The $N$-bit parallel data from the comparators are reduced to $\log_2 N$-bit parallel data by a binary encoder. Here, a 1-bit flag (which becomes high when any of the SSPD pixels absorbs a photon) is also generated by a flag generator, which ORs the outputs from the comparators. The outputs from the binary encoder and the flag generator are composed of logic-1 trains. However, the number of 1s in a train depends on the amplitude and timing of $I_{in}$, which is not convenient from the viewpoint of post-processing

[25]. Thus, each logic-1 train is converted into a single logic 1 by a rising-edge detector, which produces a logic 1 when the input changes from 0 to 1. Figure 1(c) shows a schematic of the rising-edge detector. Here, $\phi_1$ through $\phi_4$ represent excitation phases, along which logic operations are performed with a phase separation of 90°. The AND gate compares the last input with the second to last input, producing a logic 1 when the input changes from 0 to 1. The $(1+\log_2 N)$-bit parallel data from the rising-edge detectors are converted into $(1+\log_2 N)$-bit serial data by a parallel-to-serial converter [35]. Finally, $(1+\log_2 N)$-bit serial voltage signals $V_{\text{aqfp}}$, the first bit of which is a flag and the other $\log_2 N$ bits of which represent the spatial information of the SSPD array, are output from the voltage driver [36]. It should be noted that in the current design the encoder outputs a wrong pixel address when two or more SSPDs generate $I_{\text{in}}$ simultaneously. Thus, we intend to implement error detection or correction for such simultaneous firing in future design.

The RSFQ part operates as follows. The $N$ wires from the SSPD array are magnetically coupled to a stack of $N$ DC-SQUIDs, where each wire is coupled to a DC-SQUID. One of the DC-SQUIDs in the stack is turned into a voltage state when $I_{\text{in}}$ appears on one of the $N$ wires because the DC-SQUID stack is biased by a DC bias current $I_{\text{sqd}}$. As a result, the signal current $I_{\text{tmp}}$ flows into a high-sensitivity DC/SFQ converter [37], which converts $I_{\text{tmp}}$ into an SFQ-pulse train, via a resistor $R_{\text{tmp}}$. The first pulse in the SFQ-pulse train sets a voltage driver [38] to a voltage state. Note that the number of SFQ pulses in a pulse train generated by the DC/SFQ converter is not considered because only the first pulse turns the voltage driver into a voltage state, which ensures wider operating margins with regard to $I_{\text{sqd}}$ compared to a similar approach [37] (which attempts to generate a single SFQ pulse from a DC-SQUID stack). Finally, a voltage signal $V_{\text{rsfq}}$ is output from the voltage driver. The rising edge of $V_{\text{rsfq}}$ represents the temporal information of the SSPD array because the RSFQ part is composed of event-driven asynchronous circuits. Here, $V_{\text{rsfq}}$ is reset to a zero-voltage state by a reset signal generated in the AQFP part. A falling-edge detector, which is designed in a similar manner to the rising-edge detector shown in Fig. 1(c), generates a logic 1 (reset signal) when the input (flag) changes from 1 to 0, i.e., $I_{\text{in}}$ decays. Then, the reset signal is sent to the voltage driver from the falling-edge detector via an AQFP/RSFQ interface [39]. The circuit parameters of the DC-SQUID stack in the current design are as follows. The critical current of the Josephson junctions is 100 μA (the critical current density is 2.5 kA/cm$^2$ and the McCumber parameter [40] is 4), and the $LI_c$ product of a DC-SQUID is $0.386\,\Phi_0$, where $\Phi_0$ is the flux quantum. The mutual inductance between a wire from the SSPD array and the DC-SQUID is 64.8 pH, which is optimized for an $I_{\text{in}}$ of approximately 15 μA.

## 3. Experiment

As proof of concept, we demonstrate a four-pixel SSPD array using an AQFP/RSFQ hybrid interface. Figure 2 shows a micrograph of a hybrid interface for $N = 4$ that was fabricated using the 2.5 kA/cm$^2$ Nb standard process (STP2) [41] provided by the National Institute of Advanced Industrial Science and Technology (AIST). The chip die size is 5 mm by 5 mm. Note that in this design a parallel-to-serial converter is not implemented. Thus, the AQFP part has three outputs ($V_{\text{flg}}$, $V_{\text{aqfp1}}$, and $V_{\text{aqfp2}}$) and three voltage drivers, where $V_{\text{flg}}$ represents a flag, and $V_{\text{aqfp1}}$ and $V_{\text{aqfp2}}$ show the pixel address. The AQFP part is powered by a pair of AC excitation currents ($I_{x1}$ and $I_{x2}$) and a DC offset current $I_d$ [32]. In the present study, the frequency of $I_{x1}$ and $I_{x2}$ is set to 150 MHz, which corresponds to the sampling frequency of the comparators. The logical threshold of the comparators is adjusted by the reference current $I_{\text{ref}}$ [25], which sets the logical threshold of each comparator to 5 μA in this experiment. The RSFQ part is powered by the main bias current $I_b$, the bias current for the DC/SFQ converter $I_{\text{ds}}$, and the bias current for the DC-SQUID stack $I_{\text{sqd}}$. The total supply current $I_{\text{tot}}$ for the hybrid interface is 31.5 mA, most of which is attributed to $I_b$ (28.1 mA). As mentioned above, $I_{\text{tot}}$ does not increase with $N$. Thus, even for large $N$, the supply current for the hybrid interface can be kept sufficiently small for the implementation using a compact cryocooler, such as a 0.1-W GM cryocooler. Assuming that an SSPD array is biased via a single cable [42], the total cable count required for demonstrating the SSPD array using a hybrid interface is eleven, one of which is for biasing the SSPD array, two of which are for differential biasing with regard to $I_b$, and the rest of which are for $I_{ref}$, $I_{x1}$, $I_{x2}$, $I_d$, $V_{aqfp}$, $I_{sqd}$, $I_{ds}$, and $V_{rsfq}$. It is important that the cable count does not increase with $N$.

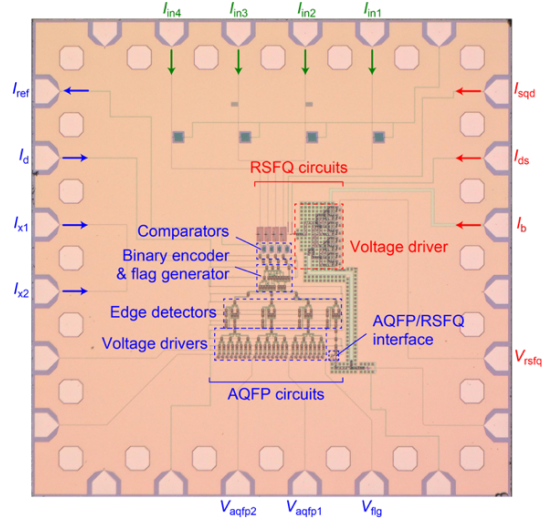

Fig. 2. Micrograph of the hybrid interface for $N = 4$. The chip die size is 5 mm by 5 mm. In this design, a parallel-to-serial converter is not implemented in the AQFP part.

Figure 3(a) shows the experimental setup for demonstrating a fiber-coupled four-pixel SSPD array using the hybrid interface shown in Fig. 2, both of which are placed in the same 0.1-W GM cryocooler. The SSPD array and the hybrid interface are stored in separate packages and are interconnected using coaxial cables and T-type connectors, as shown in Fig. 3(b), which is a photograph of the sample stage. The temperature of the sample stage was 2.49 K. The SSPD array is a four-element array [43] made of NbTiN with a total area of 40 μm × 40 μm, and each SSPD is a superconducting nanowire avalanche photon detector (SNAP) [44] in which two nanowires are connected in parallel. The switching currents of the four SSPDs range from 40 μA to 42 μA. A 10-MHz pulsed laser (Calmar Laser, FPL-02CFFNIT) applies 1,550-nm optical inputs (with a pulse width of 0.2 ps), the photon count of which ranges from 0.01 per pulse to 1,000 per pulse via a power controller and an optical attenuator, to the SSPD array. The SSPD array generates pulsed current $I_{fire}$ when detecting a photon. Half of $I_{fire}$ ($I_{in}$) is sampled by the hybrid interface, and the other half is directly output from the cryocooler. Thus, the outputs from both the SSPD array and the hybrid interface can be observed. Here, $V_{sspd1}$ through $V_{sspd4}$ represent the voltage signals from pixels 1 through 4 of the SSPD array, respectively. Moreover, $V_{flg}$ is not observed to save the number of cables. All of the outputs are amplified using low-noise amplifiers (LNAs), where $V_{sspd1}$ through $V_{ssped4}$ are amplified using LNAs (RF Bay, LNA-545) with a 500-MHz bandwidth, 45-dB gain, and 1.9-dB noise figure, and the other outputs are amplified using LNAs (RF Bay, LNA-1800) with a 1.8-GHz bandwidth, 30-dB gain, and 2.2-dB noise figure. During measurement, only one pixel is biased by dc bias current $I_{sspd}$ (i.e., the other three pixels are non-active) to easily compare the outputs from the SSPDs with those from the hybrid interface, as will be shown later. Here, $I_{sspd}$ is set to 36 μA, so the amplitude of $I_{in}$ is approximately 18 μA. For this bias condition, the detection efficiency is 5.4%, 0.024%, 0.45%, and 0.031% for pixels 1 through 4, respectively. Figures 4(a) through 4(d) show the measurement waveforms for pixels 1 through 4, respectively. In addition, $V_{sspd}$ corresponds to the voltage signal of the pixel under test. For example, $V_{sspd}$ corresponds to $V_{sspd1}$ when demonstrating pixel 1. Figure 4 indicates that the rising edge of $V_{rsfq}$ shows the timing of photon detection events and that the combination of $V_{aqfp1}$ and $V_{aqfp2}$ shows the correct pixel addresses.

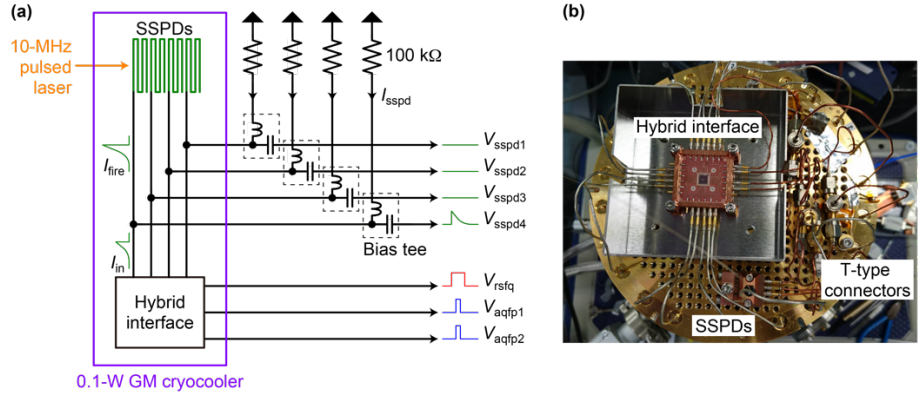

Fig. 3. (a) Experimental setup. The SSPD array and the hybrid interface are interconnected using coaxial cables in the same cryocooler. (b) Photograph of the sample stage.

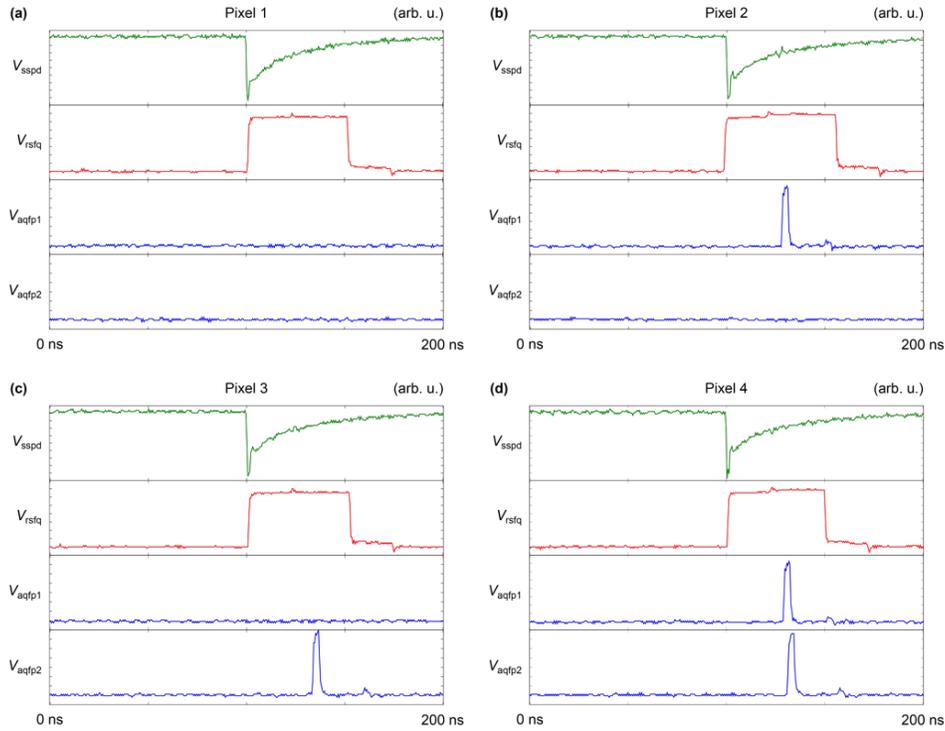

Fig. 4. Measurement waveforms for (a) pixel 1, (b) pixel 2, (c) pixel 3, and (d) pixel 4. The rising edge of $V_{rsfq}$ shows the timing of photon detection events. Here, $V_{aqfp1}$ and $V_{aqfp2}$ show the pixel addresses.

We conduct two types of experiments to demonstrate that the hybrid interface can read out the SSPD array with a low error rate and low timing jitter. First, we observe the count rates of $V_{sspd}$, $V_{rsfq}$, $V_{aqfp1}$, and $V_{aqfp2}$ using pulse counters (Stanford Research Systems, SR400) to investigate the correlation between the outputs from the SSPD array and those from the hybrid interface. Figure 5 shows the measurement results of the output count rates as a function of the laser power for each pixel. The green solid lines represent the count rates of $V_{sspd}$, and the red circles represent those of $V_{rsfq}$. The blue crosses represent the count rates of $V_{aqfp1}$, and the blue plus symbols represent those of $V_{aqfp2}$. Figure 5 indicates that both AQFP and RSFQ parts read out photon detection events with a low error rate and that correct pixel addresses are generated for all of the pixels. For instance, in Fig. 5(c) for pixel 3, the count rates of $V_{rsfq}$ and $V_{aqfp2}$ agree well with that of $V_{sspd}$, while that of $V_{aqfp1}$ is zero. This indicates that the correct pixel address (10) is generated with a low error rate. Figure 5 also shows that all of the pixels operated for single-photon events because the count rates are proportional to the laser power. Next, we measure the timing jitter of $V_{sspd}$ and $V_{rsfq}$ using an event timer (PicoQuant, HydraHarp 400). Figure 6 shows the measured timing jitter for each pixel. The green solid lines represent the timing jitter of $V_{sspd}$, and the red circles denote that of $V_{rsfq}$. Note that the photon flux is the same between the jitter measurement of $V_{sspd}$ and that of $V_{rsfq}$ because the timing jitter of $V_{sspd}$ and $V_{rsfq}$ is measured simultaneously. The full width at half maximum (FWHM) of the timing jitter of $V_{sspd}$ ranges from 71 ps to 75 ps, whereas that of $V_{rsfq}$ ranges from 66 ps to 72 ps. The FWHM of $V_{rsfq}$ is less than that of $V_{sspd}$ for all of the pixels, which indicates that the RSFQ part reads out photon detection events with low timing jitter and that the RSFQ part does not deteriorate the timing jitter of SSPDs. It is expected that the jitter of $V_{rsfq}$ can be further reduced by applying full $I_{fire}$ to the hybrid interface (i.e., $I_{in} = I_{fire}$) because the jitter of RSFQ circuits deceases as the amplitude of $I_{in}$ increases [45].

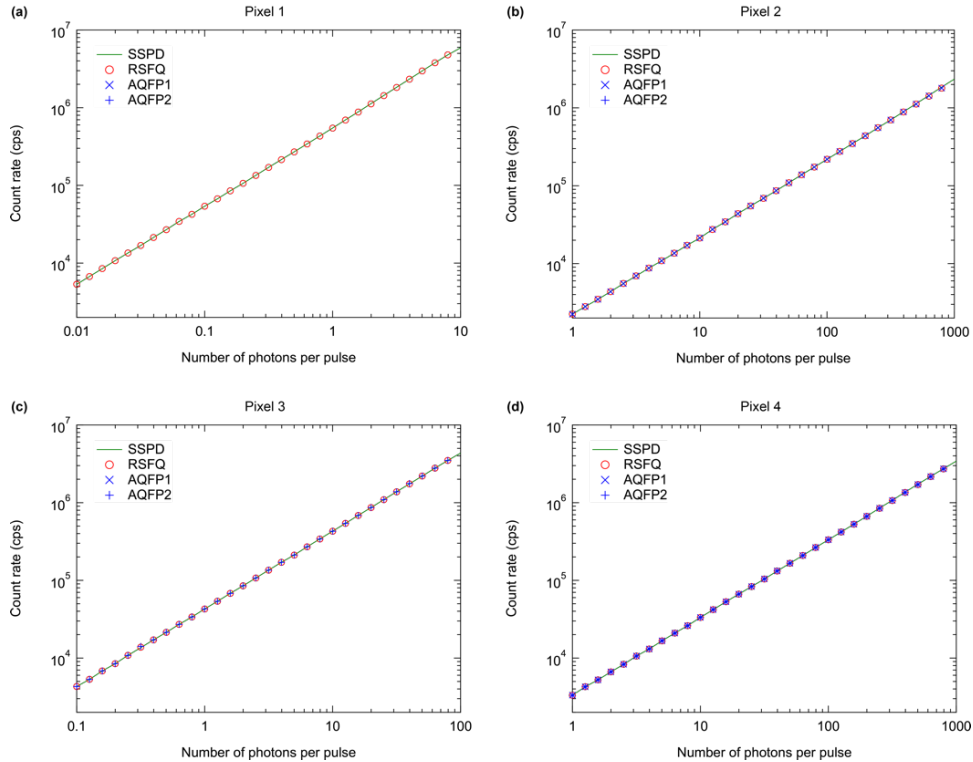

Fig. 5. Output count rates versus laser power for (a) pixel 1, (b) pixel 2, (c) pixel 3, and (d) pixel 4. The count rates for the hybrid interface correlate with the count rates for the SSPD array.

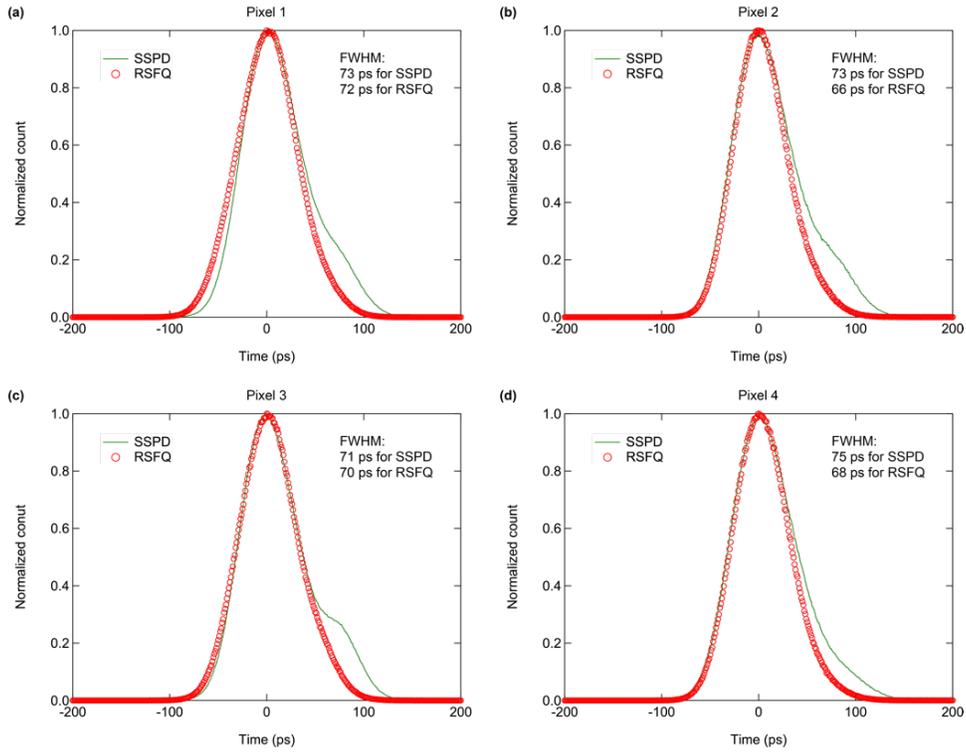

Fig. 6. Timing jitter for (a) pixel 1, (b) pixel 2, (c) pixel 3, and (d) pixel 4. The timing jitter for the hybrid interface is better than that of the SSPD array.

## 4. Summary

We proposed the AQFP/RSFQ interface as a scalable readout interface for SSPD arrays. The hybrid interface can operate with low power and low supply current, which can avoid temperature increase at the sample stage due to Joule heating caused by the supply current and parasitic resistance. We demonstrated a four-pixel SSPD array using the hybrid interface in a 0.1-W GM cryocooler. The measurement results show that the hybrid interface can read out all of the pixels with a low error rate and low timing jitter. In the future, we intend to demonstrate large SSPD arrays using the hybrid interface. Here, we discuss the maximum pixel count that a hybrid interface can handle in the light of the latency of the hybrid interface. Assuming that the latency is dominated by the duration time of $V_{aqfp}$ ($\tau_{aqfp}$) for simplicity, $\tau_{aqfp}$ limits the maximum readout frequency $f_r$. As shown in Fig. 1(a), the bit length of $V_{aqfp}$ is $1 + \log_2 N$, so that $\tau_{aqfp} = (1 + \log_2 N)/f_s$, where $f_s$ is the sampling frequency (i.e., the operating frequency of AQFP circuits). Thus, the maximum $N$ ($N_{max}$) is determined by $(1 + \log_2 N_{max})/f_s = 1/f_r$. For $f_s = 150$ MHz and $f_r = 10$ MHz, $N_{max}$ is $1.64 \times 10^4$, which can be further gained by increasing the ratio $f_s/f_r$.

## Funding

The present study was supported by KAKENHI (No. 18H05245 and No. 18H01493) from the Japan Society for the Promotion of Science (JSPS).

## Acknowledgements

The circuits were fabricated in the Clean Room for Analog-digital superconductiVITY (CRAVITY) of the National Institute of Advanced Industrial Science and Technology (AIST) using the standard process (STP2). The authors would like to thank C. J. Fourie for providing the 3D inductance extractor, InductEx.

## Disclosures

The authors declare no conflicts of interest.